\documentclass[useAMS,usenatbib]{mn2e}
\usepackage[utf8]{inputenc}
\usepackage[T1]{fontenc}
\usepackage{amsmath}
\usepackage{amssymb}
\usepackage{caption}
\usepackage{subcaption}
\usepackage[english]{babel}
\usepackage{natbib}
\usepackage{csquotes}
\usepackage{graphicx}
\graphicspath{{images/}}
\usepackage{aas_macros}
\usepackage{siunitx}

\DeclareSIUnit\erg{erg}
\DeclareSIUnit\gauss{G}

\title[A stochastic model for extreme TeV BL Lacs] 
{Extreme TeV BL Lacs: a self-consistent stochastic acceleration model} 

\author[Sciaccaluga \& Tavecchio]
{Alberto Sciaccaluga$^{1,2}$\thanks{E--mail: alberto.sciaccaluga@inaf.it, fabrizio.tavecchio@inaf.it}, Fabrizio Tavecchio$^{1}$\\
$^1$INAF -- Osservatorio Astronomico di Brera, via E. Bianchi 46, I--23807 Merate, Italy\\
$^2$Dipartimento di Fisica, Università degli studi di Genova, Via Dodecaneso 33, 16146, Italy\\
}

\begin{document}

\maketitle

\begin{abstract}
    Lately a specific kind of blazars drew the attention of the gamma-ray astronomy community: the extreme TeV BL Lacs, blazars that present an extremely energetic and hard emission at very high-energy. Explaining their features is still an open challenge, in fact the most used phenomenological models have difficulties to satisfactorily reproduce their SED. Based on a scenario we have recently proposed, we suppose that the non-thermal particles are firstly accelerated by a jet recollimation shock, which induces turbulence in the rest of the jet. Non-thermal particles are further accelerated by the turbulence, which hardens the particle spectra and accordingly the radiative emission. Given the physical properties of the plasma, as inferred by emission models, we expect a strong impact of the accelerating particles on the turbulence. Assuming isotropy and homogeneity, the interaction between non-thermal particles and turbulence and their spectra is modeled solving a system of two non-linear, coupled Fokker-Planck equations, while the radiative emission is calculated through the Synchrotron Self Compton model. The emission predicted by our model is then compared with the prototype extreme TeV BL Lac object 1ES 0229+200 and the parameters obtained to reproduce its SED are in line with the expectations.
\end{abstract}

\begin{keywords} 
    galaxies: jets --- radiation mechanisms: non-thermal ---  shock waves ---  gamma--rays: galaxies 
\end{keywords}

\section{Introduction}
\label{section:int}

Blazars are active galactic nuclei with a relativistic jet (produced in the supermassive black hole vicinity and propagating far outside the host galaxy) pointing toward the Earth \citep{romero,blandford}. Thanks to the relativistic beaming their emission is dominated by the jet non-thermal component, characterized by an emission ranging from radio to gamma-ray. The spectral energy distribution displays two broad peaks \citep{ghisellini_righi}, attributed to synchrotron and inverse Compton mechanisms (but the high-energy peak is also interpreted in terms of hadronic processes, see e.g. \citealt{bottcher_reimer}). The maximum of the second component shifts from the MeV band, for the most powerful flat spectrum radio quasars, to the TeV band for BL Lacs. Generally, the variability of the blazar emission is extreme, both in amplitude and timescale (Aharonian et al. 2006, Bonnoli et al. 2011).

Among blazars, extreme highly peaked BL Lac objects (EHBLs; \citealt{costamante,extreme_theo}) are a peculiar class which recently emerged because of their unique features. In particular we focus on the {\it extreme TeV BL Lacs}, a subclass of the EHBLs which displays extreme properties:
i) the second SED peak beyond $1\ \si{\TeV}$; 
ii) a hard sub-TeV intrinsic spectrum;
iii) in most cases, the TeV emission is stable over years.
These characteristics are hardly reproducible by a phenomenological model based only on diffusive shock acceleration by a single shock. In fact the sub-TeV spectrum hardness requires the particle spectrum to have spectral index $p<2$ (supposing $dN/d\gamma \propto \gamma^{-p}$), but with a single shock acceleration $p \geq 2$. In order to obtain harder spectra, other acceleration scenarios have been proposed, based on a hard electron distribution with a large minimum energy \citep{tavecchio_ghisellini}, a maxwellian-like electron distribution \citep{lefa_rieger}, a beam of high-energy hadrons \citep{essey_kusenko}, a lepto-hadronic model \citep{cerruti}, internal absorption \citep{aharonian_khangulyan} or emission from a large-scale jet \citep{bottcher_dermer}. 

Leptonic models which leave the acceleration process unspecified show that the mean magnetic field and, consequently, the magnetization of extreme TeV Bl Lacs are low \citep{extreme_theo}, therefore the diffusive shock acceleration should work efficiently \citep{sironi_petropoulou}. Recently \cite{zech_lemoine} proposed a model based on diffusive shock acceleration by multiple shocks. After an initial shock due to the recollimation of the jet by the external gas, several shocks could form, as confirmed by the 2D numerical simulations on jet recollimation \citep{fichet}. Non-thermal particles are then accelerated multiple times and they can reproduce the hard spectra observed for the extreme TeV Bl Lacs. However, recent 3D simulations of the jet recollimation \citep{gourgouliatos_komissarov, mhd_instability} show that if the magnetization is below a critical value, after the first shock the jet becomes unstable and turbulent, preventing the formation of additional shocks. Based on this new evidence we have proposed a different scenario: the non-thermal particles are firstly accelerated by the recollimation shock, then they are further accelerated by the turbulence formed in the downstream region. This hybrid scenario (see also \citealt{mignone}) based on diffusive shock acceleration and stochastic acceleration for extreme TeV BL Lacs has been invoked in \citealt{tavecchio_costa}, but there are some important caveats, in particular regarding the damping of the turbulence, that can modify the expected emission. Indeed, since the energy density of the non-thermal particles is much more larger than the magnetic energy density, a strong damping of the turbulence is expected. In these conditions the standard Kolmogorov spectrum adopted in \citealt{tavecchio_costa} is not self-consistent.
We note that this is a feature common to models for blazar emission based on stochastic acceleration, for which strong damping is generally expected but not self-consistently determined (see for example \citealt{kakuwa_asano}).

In this paper we therefore extend the simple treatment of \citealt{tavecchio_costa}, introducing a self-consistent model capable of describing the interplay between the turbulence damping and the non-thermal particles acceleration. 



The paper is structured as follows: the phenomenological model based on turbulence is introduced in Sect. \ref{section:model}, then the details of the numerical method are presented (Sect. \ref{section:method}) and compared with the existing data (Sect. \ref{section:results}). In Sect. \ref{section:conc} we report our conclusions. 

Throughout the paper, the following cosmological parameters are assumed: $H_0=70{\rm\;km\;s}^{-1}{\rm\; Mpc}^{-1}$, $\Omega_{\rm M}=0.3$, $\Omega_{\Lambda}=0.7$.

\section{The model}
\label{section:model}

Along the lines of the scenario proposed in \citealt{tavecchio_costa}, the model presented in this paper is based on a double acceleration mechanism, an hybrid of diffusive shock acceleration and stochastic acceleration due to turbulence (see also Kundu et al. 2020). 

We suppose that the non-thermal particles are first accelerated by a recollimation shock, formed in the jet when the external pressure is greater than the jet pressure. As already explained, the shock acceleration process is a deterministic process and the output spectra of the non-thermal particles is a power law, with spectral index $p\geq 2$ \citep{blandford_eichler, sironi_spitkovsky}. 


The non-thermal particles injected by the recollimation shock can be further energized interacting with the turbulence, hardening the distribution and therefore the emitted spectrum. Non-thermal particles are supposed to be accelerated mostly through gyroresonant interaction with the weakly turbulent Alfvén waves propagating parallel or antiparallel to the mean magnetic field, therefore the phase speed is equal to the Alfvén speed (for non-resonant acceleration see \citealt{bresci_lemoine}). This acceleration process is stochastic and in order to study the equilibrium spectra of both the non-thermal particle spectra and the turbulence it is necessary to consider the time evolution of their phase-space densities through two momentum diffusion equations.

Assuming isotropy and homogeneity, the momentum diffusion equation can be highly simplified, obtaining the classical Fokker-Planck equation. Therefore our goal is to solve a system of two coupled non-linear Fokker-Planck equations, which describe both the interaction and the spectra of the non-thermal particles and of turbulence. The emission recorded by the observer is produced in the shock downstream region and comprises non-thermal particles at different stages of acceleration, from the shock to the top of the emitting region, which is modelled as a cylinder of radius $R$ and length $10 \,R$. The length of the emission region, where the instability develops and injects turbulence in the flowing plasma, is roughly based on the simulations shown in \citealt{mhd_instability}. We can be relatively confident that within such a distance, the magnetic field decay and the adiabatic losses effectively quench the emission, see also \citealt{tavecchio_costa}. All quantities appearing in equations below are expressed in the downstream jet frame, we will move to the observer frame only for the quantities linked to the radiative emission. 

The first equation describes the time evolution of $f(p)$, the momentum distribution of the non-thermal particles, from here on supposed to be relativistic electrons, since we are considering a leptonic model:
\begin{equation}
    \frac{\partial f}{\partial t} = \frac{1}{p^2} \frac{\partial}{\partial p} \left [ p^2 D_p \frac{\partial f}{\partial p} + p^2 \left ( \frac{\partial p}{\partial t} \right )_{\text{rad}} f \right ] + \frac{f}{t_\text{esc}} +I_f
    \label{eq:fp_part}
\end{equation}
where $p$ is the particle momentum, $D_p$ the momentum diffusion coefficient, $\left (\partial p/\partial t \right )_{\text{rad}}$ the cooling coefficient due to radiative emission and $I_f$ is the particle injection rate. Starting from momentum distribution, it is possible to calculate the electron energy density, in fact $n(\gamma) = 4 \pi p^2 f(p) m_e c$, where $m_e$ is the electron mass. Supposing only parallel and antiparallel propagating MHD waves interacting with electrons, the acceleration time can be estimated as \citep{miller_roberts,kakuwa}:
\begin{equation}
    t_\text{acc} = \left[ \frac{2 \beta_w^2 c}{U_B r_g}\int_{k>k_\text{res}} \frac{W_B(k)}{k} dk \right]^{-1}
    \label{eq:acc_time}
\end{equation}
where $\beta_w = v_a/c$ is the wave velocity in unit of c, $U_B = B^2/8\pi$ is the magnetic energy density, $r_g = \gamma m_e c^2/eB$ is the Larmor radius, $W_B(k) \approx W(k)/2$ is the magnetic component of the energy density of turbulence fields per unit of $k$ (indicated with $W(k)$) and $k_\text{res} = 2 \pi/r_g$ is the resonant wavenumber. Note that an electron with a fixed $\gamma$ can interact only with Alfvén waves whose wavelength is smaller than its Larmor radius.  From the time acceleration the momentum diffusion coefficient can be obtained using $D_p = p^2/2t_\text{acc}$. Neglecting inverse Compton and adiabatic cooling for simplicity, the cooling time is equal to:
\begin{equation}
    t_\text{cool} = \frac{6 \pi m_e c}{\sigma_T B^2 \gamma}
\end{equation}
where $B$ is the mean magnetic field and $\sigma_T$ is the Thomson cross section. From the cooling time it is possible to  calculate the cooling coefficient, in fact $t_\text{cool} = p/\dot{p}$. Another effect influencing the electron spectra is the spatial escape from the acceleration site:
\begin{equation}
    t_\text{esc} = \frac{R}{c} + \frac{R^2}{k_\parallel}
    \label{eq:escape_time}
\end{equation}
where $k_\parallel = c r_g /9 \zeta(k_\text{res})$ is the spatial diffusion coefficient along the mean magnetic field, while $\zeta(k) = kW_B(k)/U_B$ is the relative amplitude of the turbulent magnetic field energy density for a given $k$. Note that if the mean free path is large (which implies a small diffusion coefficient), the escape time is simply equal to the geometric escape time, that is the ratio between the dimension of the acceleration region and the particles velocity. Finally the last term of the equation \eqref{eq:fp_part} describes the particle injection, identified here as the electrons accelerated at the recollimation shock and advected downstream. Supposing a strong shock, the injected electron density per unit time is equal to:
\begin{equation}
    I_n(\gamma) = I_{n,0} \,\gamma^{-2}\, e^{-\frac{\gamma}{\gamma_\text{cut}}} \quad  \text{with} \quad  \gamma_\text{inj}<\gamma<100\,\gamma_\text{cut}
\end{equation}
The extremes of the injection range are fixed and they are estimated thanks to recent simulations of diffusive shock acceleration: $\gamma_\text{inj} = 10^3$ and $\gamma_\text{cut} = 10^5$ \citep{zech_lemoine}. From the injected electron density per unit time it is possible to obtain the injection distribution in the momentum space, in fact $I_n(\gamma) = 4\pi p^2 m_e c I_f(p)$.

The second equation we have to consider describes in a phenomenological way the time evolution of $W(k)$, the wavenumber energy distribution of the turbulence, i.e. the energy density of waves with wavenumber between $k$ and $k+dk$. Defining $Z(k) = W(k)/k^2$, in order to obtain a Fokker-Planck equation in its canonical form, the equation can be written as \citep{eilek, zhou_matthaeus, kakuwa}: 
\begin{equation}
        \frac{\partial Z}{\partial t} = \frac{1}{k^2} \frac{\partial}{\partial k} \left ( k^2 D_k \frac{\partial Z}{\partial k}  \right ) + \frac{Z}{t_\text{dam}} + \frac{I_W}{k^2}   
        \label{eq:fp_turbu}
\end{equation}
where $D_k$ is the wavenumber diffusion coefficient, $t_\text{dam}$ the damping time and $I_W$ the turbulence injection rate. We neglect spatial effects, such as wave propagation outside the downstream region, and dissipation mechanisms, such as Landau or cyclotron damping. Moreover there is an important caveat: equation \eqref{eq:fp_turbu} describes an isotropic turbulent field, while the acceleration time formula is calculated assuming that turbulence consists exclusively in waves parallel or antiparallel to the magnetic field. Even though it would be more correct to recompute equation \eqref{eq:acc_time} in the case of isotropic turbulence, we adopt the convention used in the literature. 

The wavenumber diffusion coefficient depends on which scenario we are considering. Since we suppose that the jet magnetization is weak and the turbulence is strong, the Kolmogorov phenomenology must be applied \citep{zhou_matthaeus, miller_roberts}, which implies:
\begin{equation}
    t_\text{cas} = \frac{1}{k \beta_w c} \sqrt{\frac{2U_B}{kW(k)}}
    \label{eq:cascade_time}
\end{equation}
The damping time, instead, can be calculated from the acceleration term in equation \eqref{eq:fp_part}, requiring energy conservation (for further details see \citealt{miller_roberts, kakuwa}). In fact the energy required to accelerate and diffuse the electrons in the momentum space cannot be assumed infinite, since in the considered scenario the jet magnetization is weak and then the turbulence is extremely damped:
\begin{equation}
   t_\text{dam} = \left | \frac{4\pi e^2 \beta_w^2 }{m_e c k} \int_{\gamma>\gamma_\text{res}} \gamma^2 \frac{\partial}{\partial \gamma} \left ( \frac{n(\gamma)}{\gamma^2} \right ) d\gamma \right |^{-1}
\end{equation}
where $\gamma_\text{res}$ is the resonant Lorentz factor, defined by $k = 2 \pi/r_g(\gamma_\text{res})$. In fact an Alfvén wave with a fixed wavenumber can interact only with electrons whose Larmor radius is greater than its wavelength. 

Finally the last term in equation \eqref{eq:fp_turbu} represents the turbulence injection. We suppose that the dimension of the initial turbulence wavelength is in the order of $L=R/10$:
\begin{equation}
    I_W(k) = I_{W,0}\, \delta (k-k_0)
\end{equation}
where $\delta$ is the Dirac function and $k_0 = 2\pi/L$. The radiative emission is calculated through the Synchrotron Self Compton model (see \citealt{ghisellini_chiaberge} for the synchrotron emission and \citealt{blumenthal_gould} for the inverse Compton emission), while for the beaming we use the standard formula (see the discussion in \citealt{zech_lemoine}). 

\section{Numerical method}
\label{section:method}

For solving the two coupled non linear Fokker-Planck equations, we decided to use the robust implicit Chang-Cooper algorithm \citep{chang_cooper}, in which the diffusion and the cooling terms are regrouped in a single flux term, whereas for the injection and the escape/damping terms we follow the technique already implemented in other papers \citep{park_petrosian, ghisellini_chiaberge}. 

The time mesh is equally spaced and consists of $500$ points, while momentum and wavenumber are divided in two logarithmic equally spaced meshes of $400$ points: the momentum mesh starts from the momentum corresponding to $\gamma_\text{min} = 10^1$ and ends when $\gamma_\text{max} = 10^9$, whereas the wavenumber space ranges from the injection wavenumber ($k_\text{min} = k_0$) to one thousand times the maximum resonant wavenumber ($k_\text{max} = 1000\, k_{\text{res},\text{max}}$). Since the momentum and the wavenumber meshes are logarithmic, we substitute both variables with the corresponding logarithms, so both meshes result equally spaced with respect to the new variables \citep{miller_roberts}. For the same reason, the coefficient at midpoints are evaluated through the geometric mean, instead of the usual arithmetic mean \citep{park_petrosian}. 

The sizes of the Lorentz factor and wavenumber spaces are strongly linked to the boundary conditions we imposed. The extremes of the Lorentz factor space are chosen to be far from the range responsible for the radiative emission, where the particle are the most, therefore we set $f(p_\text{min}) = f(p_\text{max}) = 0$. Considering the wavenumber space, we set the flux through the lower limit to zero ($F_W(k_\text{min}) = 0$), because the turbulence energy can not spread to wavenumber corresponding to larger vortex, while the upper limit condition is less obvious, in fact both setting the flux or the function itself to zero are unphysical conditions \citep{park_petrosian}. For this reason we choose to extend the upper limit of the wavenumber space beyond the maximum resonant wavenumber. In fact between $k_{\text{res},\text{max}}$ and $k_\text{max}$ there are neither source term nor damping term, therefore in this interval we have $W(k) \propto k^{-5/3}$ and it is possible to obtain the value of the turbulence spectrum in $k_\text{max}$ simply extending the function. 

For the initial conditions, we suppose that at start there are no non-thermal particles in the downstream region ($f(t_0,p) = 0$), while for the turbulence spectrum we presume that the system is in equilibrium ($W(t_0,k) \propto k^{-5/3}$) and the normalization of the initial spectrum is supposed to be much smaller than the injection one. A no null initial condition is necessary for the injection to spread to higher wavenumbers, as it is clear from equation \eqref{eq:cascade_time}. 

Another problem for the system resolution is the non linearity of equation \eqref{eq:fp_turbu}, due to the presence of $W(k)$ in cascade time definition. Considering equation \eqref{eq:cascade_time}, there are two possible solutions: the semi-implicit method, in which we substitute $W(k)$ with the spectrum obtained in the previous time step, or the fully implicit method, where $W(k)$ is set equal to the spectrum we want to obtain in the considered time step. The first technique requires to solve a linear system, but several time steps are necessary to converge, on the other hand the second method converges despite of the width of the time step, but it requires the resolution of a non linear system. We opt for the second method, but we use an iterative procedure that requires solving several linear systems: firstly we set $W(k)$ equal to the result obtained at the previous time step, then we apply Chang Cooper algorithm, we substitute $W(k)$ with the solution, we reapply the algorithm and compare the two results. This process continues until a defined convergence criterion is satisfied \citep{larsen_levermore}: in our case we require the relative error of $W(k)$ to be smaller than $10^{-3}$.    

\section{Results}
\label{section:results}

With the setup described above, our model depends on five free parameters: the downstream region radius $R$, the Alfvén velocity $v_a$, the mean magnetic field $B$, the injected powers of the electrons and the turbulence in the jet frame, respectively $P_n^\prime$ and $P_W^\prime$. The latter two quantities are defined as follows:
\begin{gather}
    P_n^\prime = V \int \gamma m_e c^2 I_n(\gamma) \,d\gamma \\
    P_W^\prime = V \int I_W(k) \,dk 
\end{gather}
where $V = 10 \pi R^3$ is the volume of the emitting region. From these definitions we can calculate both the injection normalizations $I_{n,0}$ and $I_{W,0}$.

We apply our model to the prototypical extreme TeV BL Lac object, i.e. 1ES 0229+200 ($z = 0.1396$). 
\begin{figure}
    \centering
    \includegraphics[width = \columnwidth]{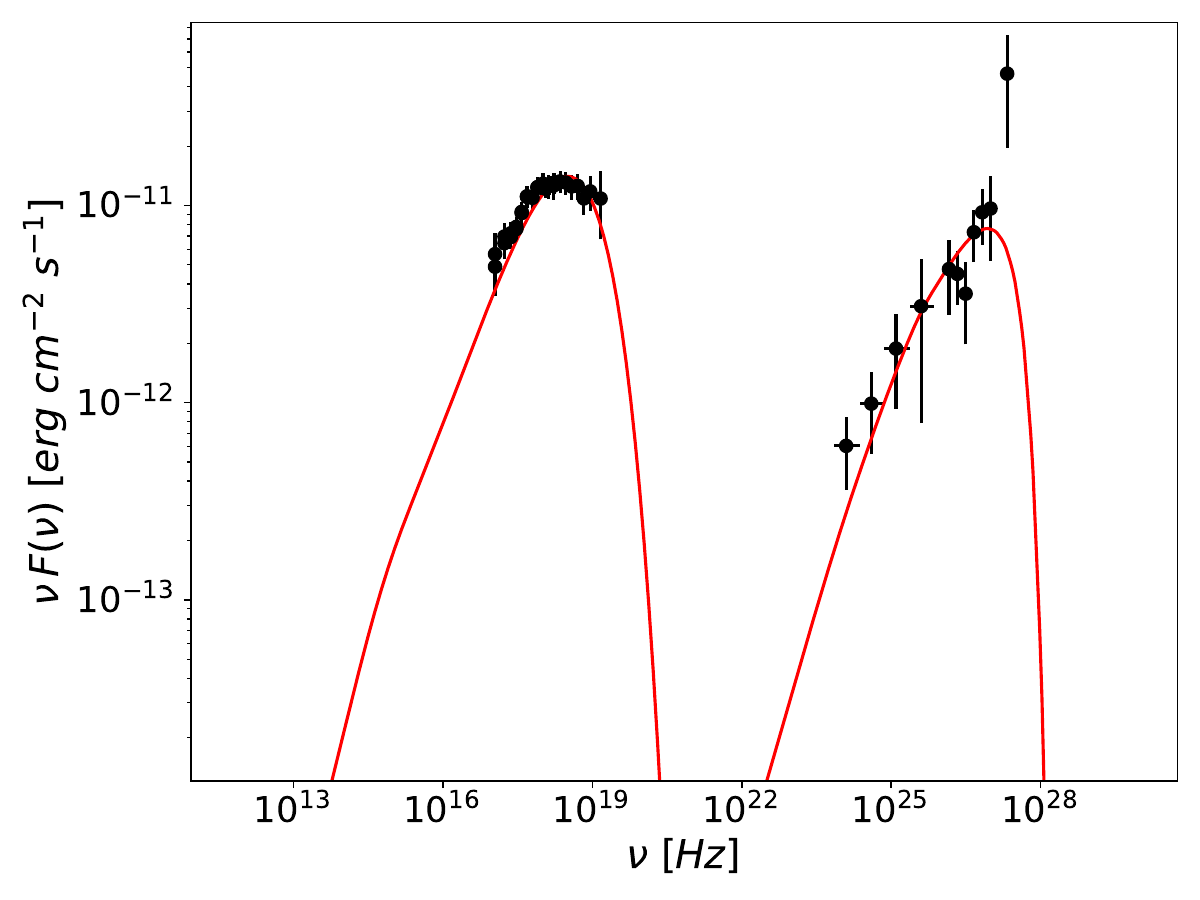}
    \caption{SED of 1ES 0229+200 (points with error bars, from Costamante et al. 2018) reproduced with the model presented in the paper (red solid line)}
    \label{fig:flux}
\end{figure}
Fig. \ref{fig:flux} shows a model realization ($R = 1.2 \times 10^{16} \,\si{\cm}$, $v_a = 2 \times 10^9\,\si{\cm/\s}$, $B = 15.9\,\si{\milli \gauss}$, $P_n^\prime = 7 \times 10^{39}\, \si{\erg/\s}$, $P_W^\prime = 7\times 10^{39}\, \si{\erg/\s}$). In this first illustration of the model, the relativistic Doppler factor is fixed for simplicity to $\delta = 25$. In applying the model to a full sample of sources, the relativistic Doppler factor could be included in the set of the free parameters. The realization shown in Fig \ref{fig:flux} is in good agreement with the data: in fact the X-ray peak is well reproduced and the model realization presents a hard sub TeV spectrum with a gamma peak beyond $1\,\si{\TeV}$. The free parameters are consistent with what we expect from the theory. Since the emission is due to the post-reconfinement flow, $R$ is much smaller than the ordinary jet radius \citep{bodo_tavecchio}, whereas the Alfvén velocity is large with respect to the usual values in order to make the stochastic acceleration energy gain relevant for the non-thermal particles. Moreover the magnetic field is in the order of the values obtained by the phenomenological models that leave the acceleration process unspecified \citep{extreme_theo}. Finally the injected power of the non-thermal particle in the jet frame is in agreement with the theory predictions \citep{ghisellini_tavecchio}.
To check the consistency of the model realization shown in Figure \ref{fig:flux} further, we also calculate some derived quantities, i.e. the jet magnetization $\sigma = 4.4 \times 10^{-3}$, the integrated relative amplitude of the turbulent magnetic field $\zeta = 0.1$, the background plasma number density $n_p = 2.8 \ \si{\cm^{-3}}$ and the final electron number density $n_e = 2.6 \times 10^{-2} \ \si{\cm^{-3}}$: they all assume satisfactory values. It should be noted that in the paper by \citealt{mhd_instability} the authors estimate the critical magnetization for typical pc-scale AGN jets to be about $\sigma_{cr} \sim 2.0 \times 10^{-3}$. Given the uncertainty of the (heuristic) estimate proposed by Matsumoto and the simplicity of our model, we are confident that the (small) difference is negligible. In any case, in Costa et al. in prep. we will try to better understand the connection between the jet magnetization and the development of turbulence, with jet parameters tailored on the extreme blazar case.  
\begin{figure}
    \centering
    \includegraphics[width = \columnwidth]{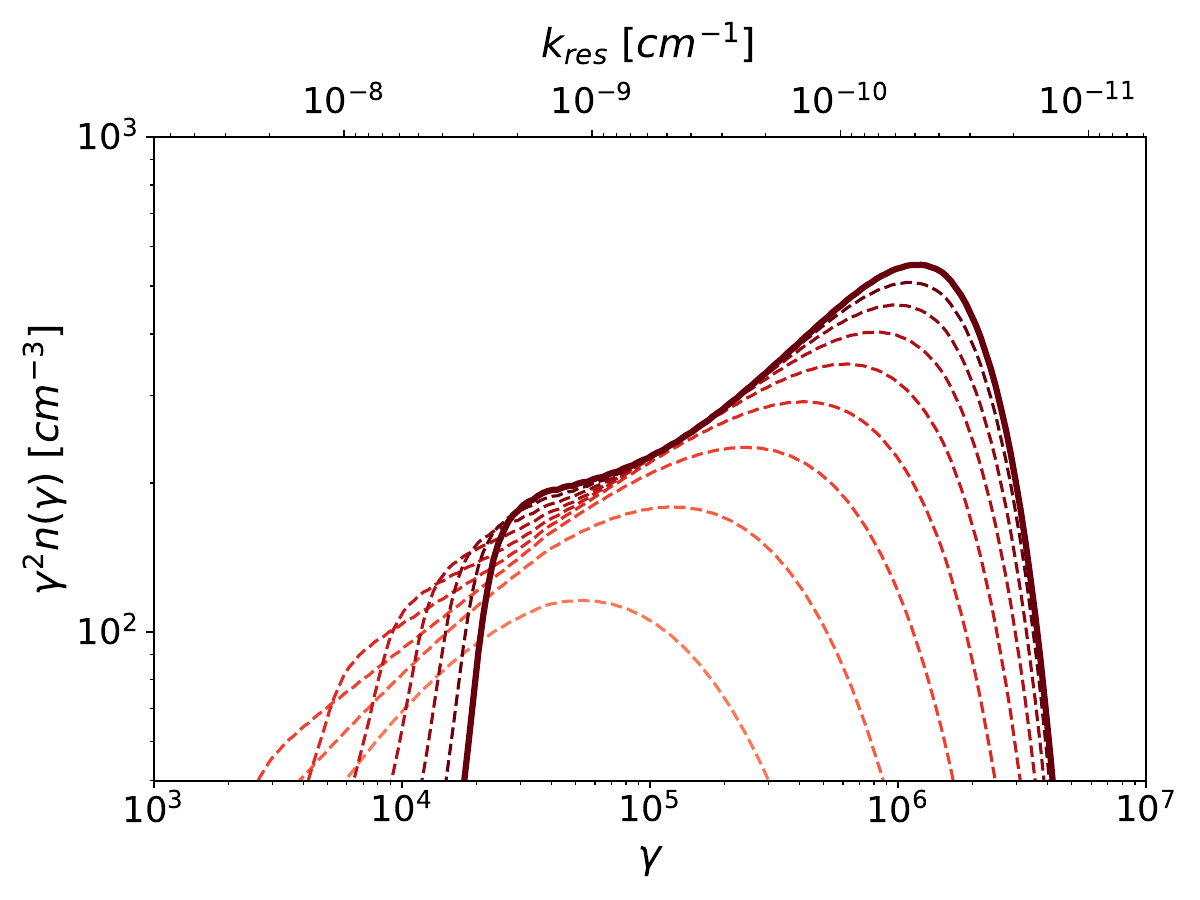}
    \caption{The time evolution of the electron energy distribution used to produce Figure \ref{fig:flux}. The dashed lines represent different time steps (each equal to $R/c$), while the solid line is the final distribution (reached at $t=10 R/c$). The time evolution is represented thanks to the red gradient. The injection is not visible.}
    \label{fig:electrons}
\end{figure}

\begin{figure}
    \centering
    \includegraphics[width = \columnwidth]{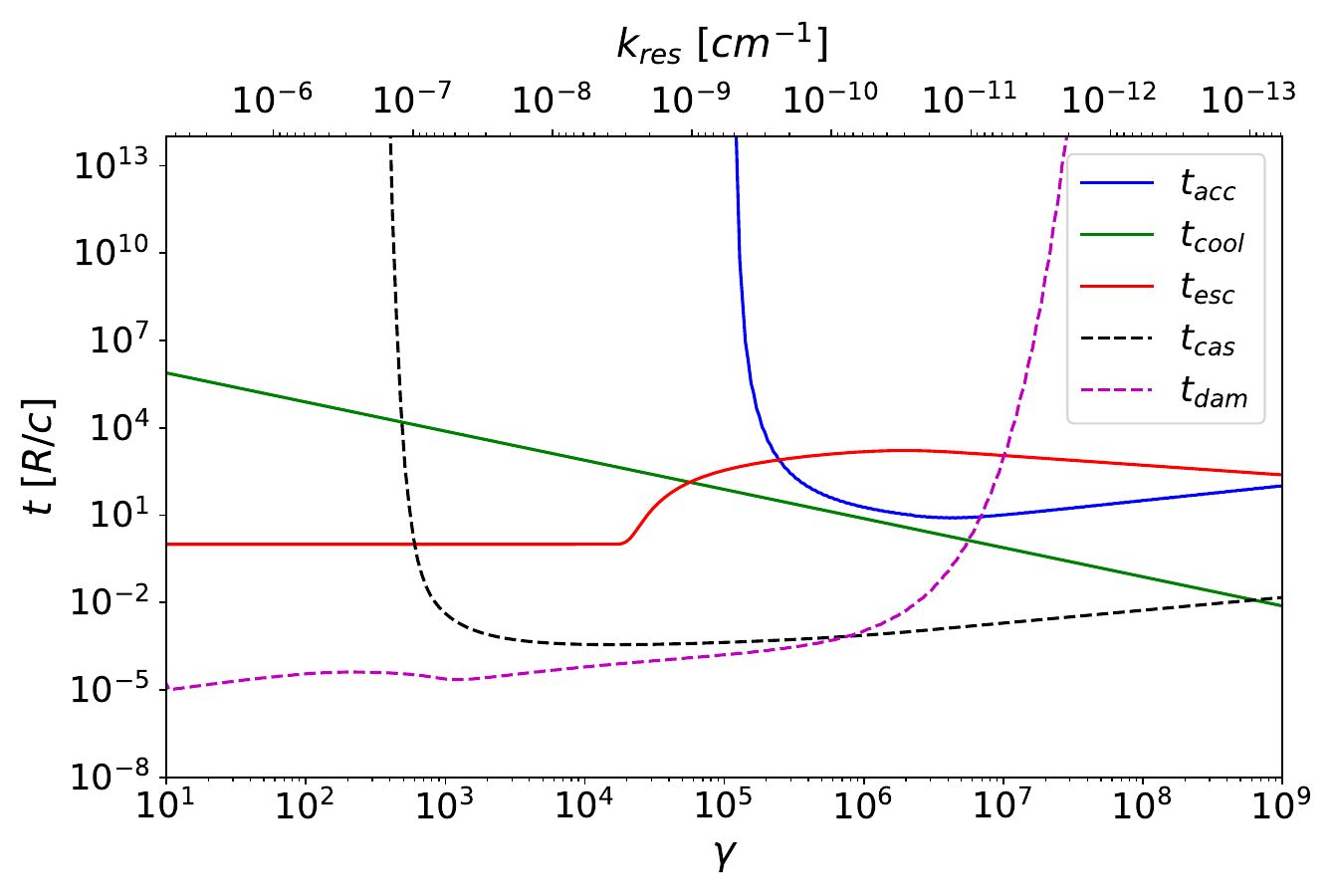}
    \caption{Timescales characterizing the system at $t=10R/c$. Times related to non-thermal particles are represented using solid lines, while for timescales for turbulence we use dashed lines.}
    \label{fig:times}
\end{figure}

In Fig. \ref{fig:electrons} the time evolution of the electron energy distribution is represented and several features can be noted. Firstly the maximum energy of the distribution increases with time, reaching after $10\, R/c$ the equilibrium value at $\gamma \sim 10^6$, where the acceleration and cooling time are nearly equal (see Figure \ref{fig:times}). Beyond the peak the final spectrum presents a steep cut-off due to the synchrotron cooling, in fact in this range $t_\text{cool} \ll t_\text{acc}$. Furthermore after some time the distribution plunges for Lorentz factor smaller than $\gamma \sim 10^4$, because for these particles the turbulence becomes negligible, then the escape time is simply equal to $R/c$, as it is clear looking to the escape time in Fig. \ref{fig:times}.     
\begin{figure}
    \centering
    \includegraphics[width = \columnwidth]{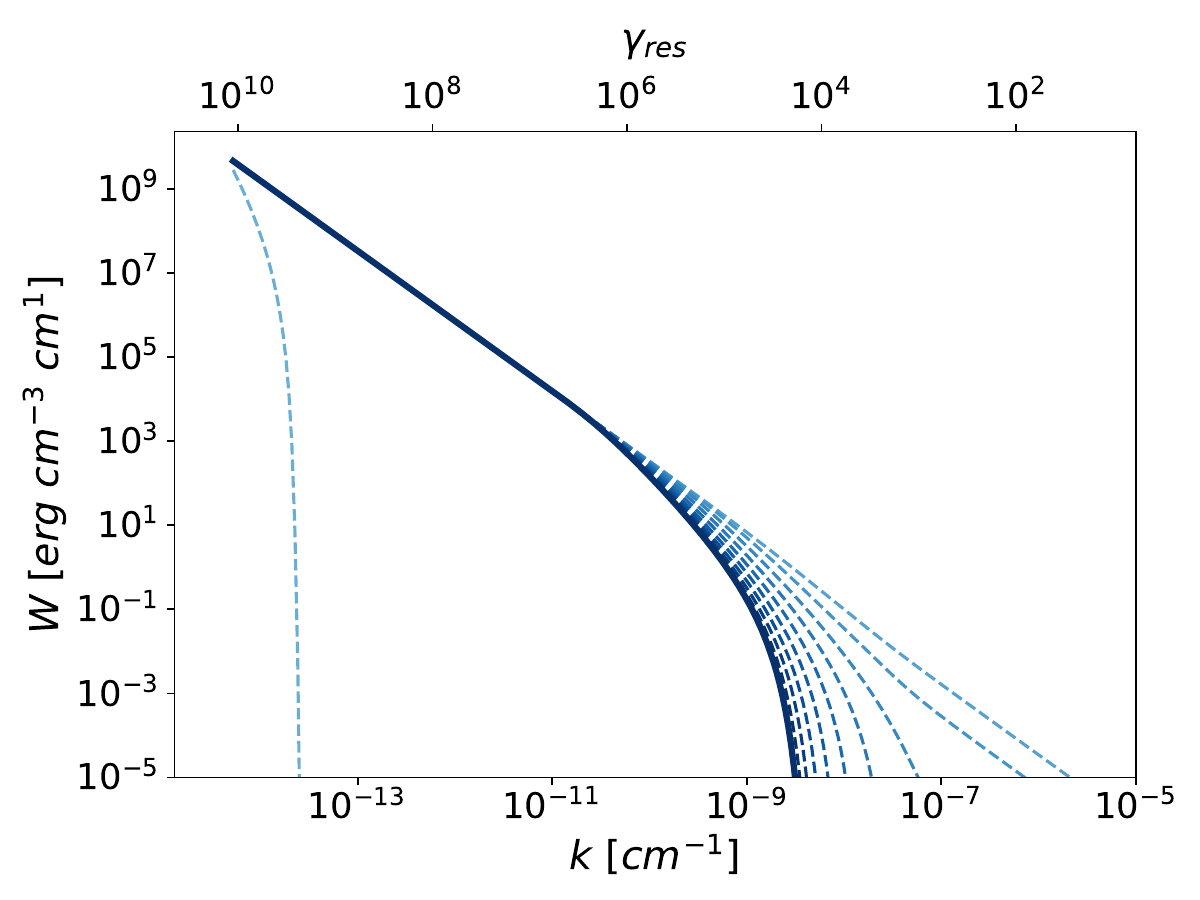}
    \caption{The time evolution of the turbulence spectrum used to produce Figure \ref{fig:flux}. The dashed lines represent different time steps (each equal to $R/c$), while the solid line is the final distribution. The time evolution is represented thanks to the blue gradient.}
    \label{fig:turbulence}
\end{figure}

Considering the time evolution of the turbulence spectrum in Fig. \ref{fig:turbulence}, initially the injection cascades to larger wavenumbers, forming the usual power law spectra, i.e. $W(k) \propto k^{-5/3}$. After spreading in all wavenumber space, the non-thermal particles start to absorb the turbulence energy. For this reason at wavenumbers larger than $k \sim 10^{-10}\,\si{\cm^{-1}}$, i.e. when $t_\text{dam} \ll t_\text{cas}$ (see Fig.\ref{fig:times}), the final turbulence spectrum is extremely damped, while at shorter wavenumbers it remains a power law, in fact in this range $t_\text{cas} \ll t_\text{dam}$. 

In order to better understand the role of the damping in shaping the electron energy distribution and the corresponding produced SED, we decide to compare our model with the data of 1ES0229+200, but excluding the damping term. 
\begin{figure}
    \centering
    \includegraphics[width = \columnwidth]{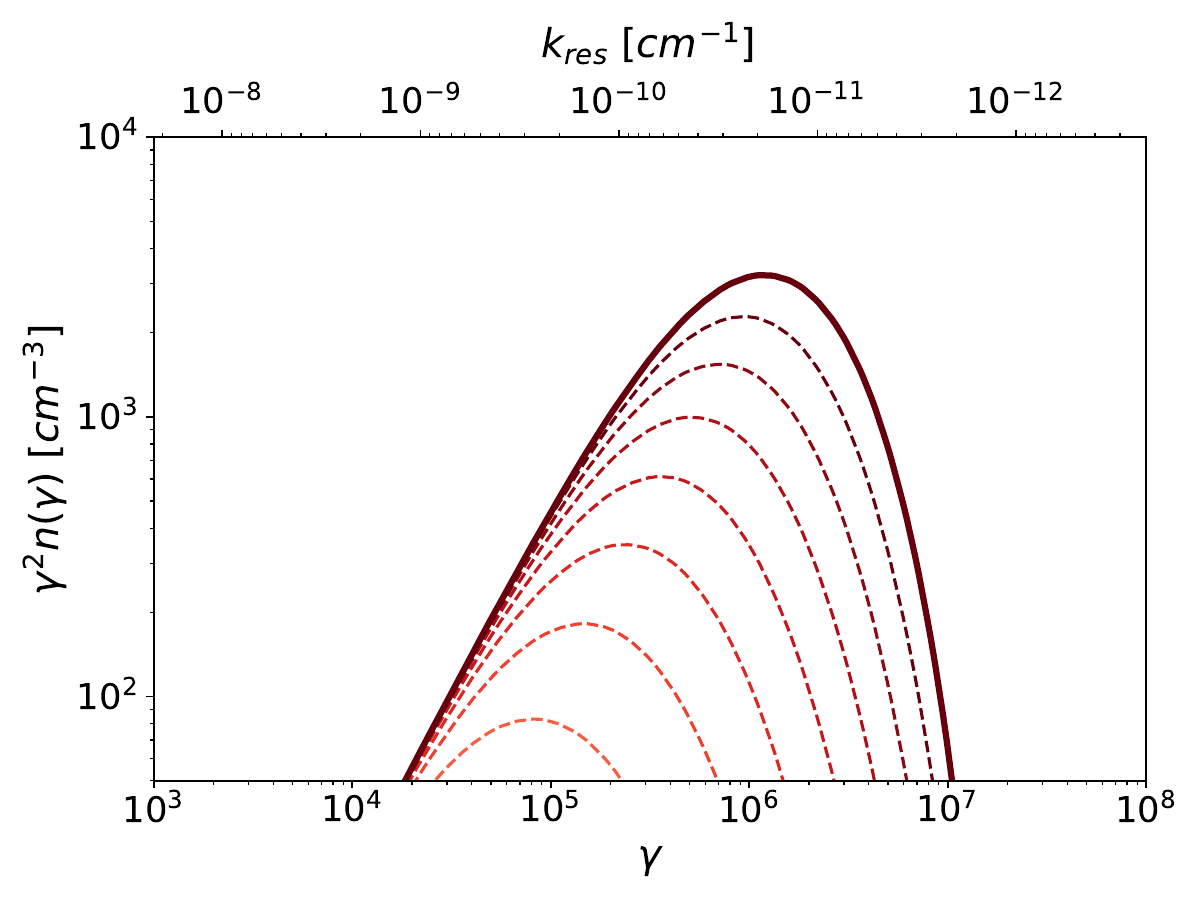}
    \caption{The time evolution of the electron energy distribution used to produce Figure \ref{fig:flux_no_damp}. The dashed lines represent different time steps (each equal to $R/c$), while the solid line is the final distribution (reached at $t=10 R/c$). The time evolution is represented thanks to the red gradient. The injection is not visible.}
    \label{fig:electrons_no_damp}
\end{figure}

The absence of damping implies that a large amount of energy is stored in waves at large wavenumbers ($k>10^{-10}\, \si{cm^{-1}}$, above the cut-off resulting from damping visible in Fig. \ref{fig:turbulence}), causing the efficient acceleration of particles at energies $\gamma\lesssim 10^5$. The rapid acceleration, in turn, results in an electron energy distribution quite hard before the peak (compare Figs. \ref{fig:electrons} and \ref{fig:electrons_no_damp}). Instead near and beyond the peak the electron spectrum remains substantially unchanged: in fact even with damping, in the corresponding wavenumber range the turbulence spectrum is the standard power law.  

\begin{figure}
    \centering
    \includegraphics[width = \columnwidth]{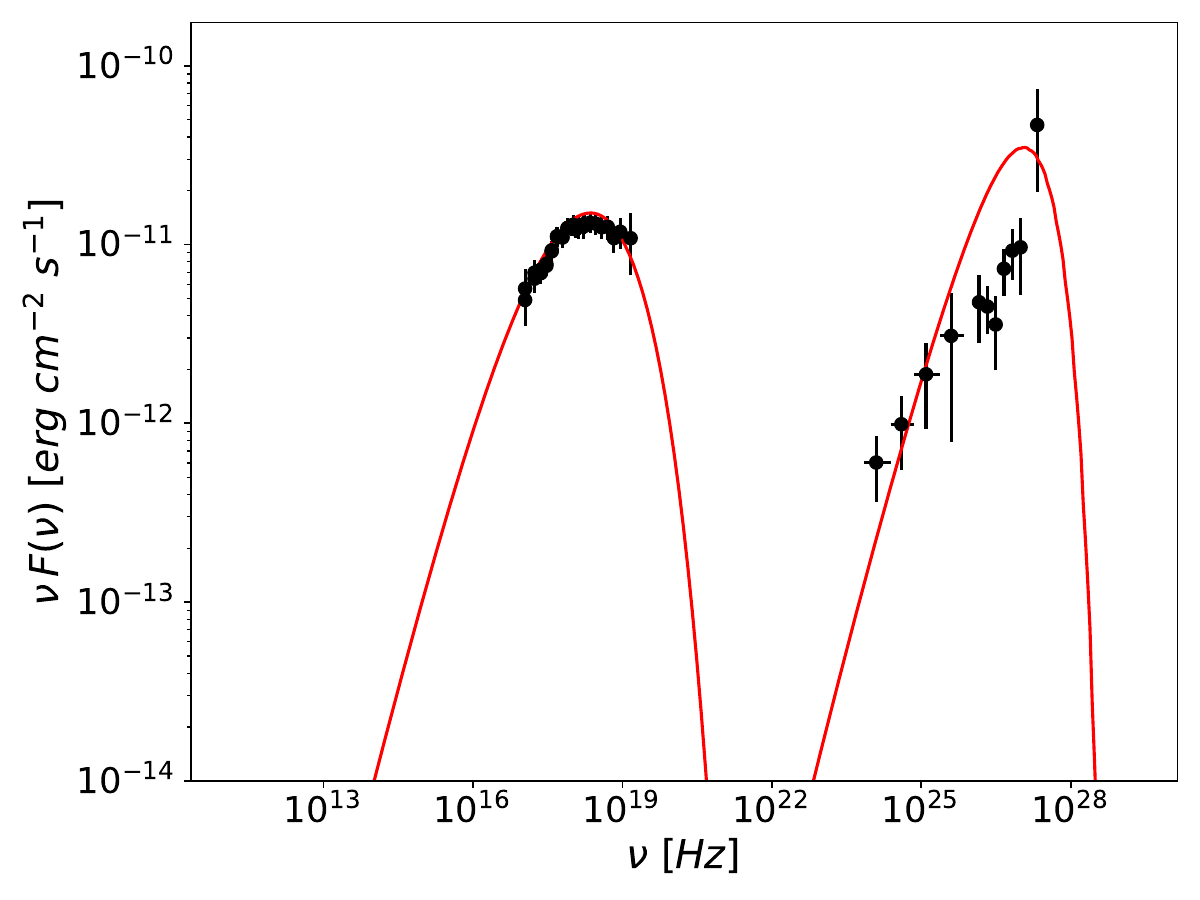}
    \caption{SED of 1ES 0229+200 (points with error bars, from Costamante et al. 2018) reproduced with the model presented in the paper (red solid line) excluding damping. Note the hard SSC spectrum.}
    \label{fig:flux_no_damp}
\end{figure}

The very narrow, maxwellian-like electron distribution translates into a very narrow and hard synchrotron and SSC peaks in the SED. As shown in Fig. \ref{fig:flux_no_damp}, while the first peak is well reproduced, the rise of the SSC component in the sub-TeV range is too hard with respect to the data, as already noted in \cite{tavecchio_costa}. Therefore, the inclusion of the damping, besides making the model self-consistent, results in a softer gamma-ray spectrum, in better agreement with the data.

\section{Conclusions}
\label{section:conc}

In this paper we have presented a self-consistent model for extreme TeV BL Lac objects, improving the simple version of \cite{tavecchio_costa}. In particular, the inclusion of damping for the turbulence results in an electron distribution with an excess of particles below the peak, around $\gamma \simeq 10^4-10^5$, which produces a softer SSC component in the VHE band, in better agreement with the data with respect to the simple model without wave damping.

Our model is able to reproduce the SED of the prototypical extreme TeV BL Lac object 1ES 0229+200, but there are several caveats. First of all it will be necessary to compare our model with the data of other extreme TeV BL Lacs in order to check if this hybrid emission scenario can be generalized to the entire population. Since the classification of this type of sources is still unclear, a different emission scenario could trigger a new way to categorize these objects. Moreover the parameters obtained by the comparison between SEDs and the model have to be checked with the results of MHD simulation. The papers already published shows that the low magnetization scenario favors the development of turbulence after the recollimation shock instead of the formation of multiple shocks. A better characterization of the emission region and turbulence based on dedicated MHD simulations will be presented in the near future (Costa et al., in prep). 

Considering the model in this paper, there are some possible improvements. Firstly the Inverse Compton (i.e. SSC) cooling term could be added in the diffusion equation of non-thermal electrons, which would make the equation \eqref{eq:fp_part} non-linear too: if necessary, the same solution proposed for the non-linearity of the diffusion equation of turbulence could be used. Moreover the computational scheme used to solve the system in this paper could be replaced with more complex and accurate algorithms, see for example \citealt{mignone} for an approach based on the employment of Runge-Kutta Implicit-Explicit schemes. 

In the following years the number of extreme TeV BL Lacs will probably increase and their SED measurements will improve. In fact many data are going to be collected, since these sources are already studied by several telescopes \citep{extreme_nustar, extreme_magic} and they will also be objects of interest for future gamma-ray facilities, such as CTA, because of their extreme high energy emission. Finally, another observational feature useful to test our model will be the polarization properties of the extreme TeV BL Lacs. Polarimetric measures in the optical band are extremely difficult for this type of sources because of the weak emission of the jet in this band, dominated by the emission of the host galaxy. Instead in the X-ray band the emission is mainly due to the synchrotron cooling of the jet particles emitting in a highly turbulent field, therefore we expect a low degree of polarization. This prediction could be proved by the IXPE mission \citep{ixpe}, the first satellite able to measure the polarization in the X-ray band, launched at the end of 2021. 

\section*{Data availability}

Data available on request.

\bibliographystyle{mnras}
\bibliography{bibliography.bib}

\begin{thebibliography}{}
\makeatletter
\relax
\def\mn@urlcharsother{\let\do\@makeother \do\$\do\&\do\#\do\^\do\_\do\%\do\~}
\def\mn@doi{\begingroup\mn@urlcharsother \@ifnextchar [ {\mn@doi@}
  {\mn@doi@[]}}
\def\mn@doi@[#1]#2{\def\@tempa{#1}\ifx\@tempa\@empty \href
  {http://dx.doi.org/#2} {doi:#2}\else \href {http://dx.doi.org/#2} {#1}\fi
  \endgroup}
\def\mn@eprint#1#2{\mn@eprint@#1:#2::\@nil}
\def\mn@eprint@arXiv#1{\href {http://arxiv.org/abs/#1} {{\tt arXiv:#1}}}
\def\mn@eprint@dblp#1{\href {http://dblp.uni-trier.de/rec/bibtex/#1.xml}
  {dblp:#1}}
\def\mn@eprint@#1:#2:#3:#4\@nil{\def\@tempa {#1}\def\@tempb {#2}\def\@tempc
  {#3}\ifx \@tempc \@empty \let \@tempc \@tempb \let \@tempb \@tempa \fi \ifx
  \@tempb \@empty \def\@tempb {arXiv}\fi \@ifundefined
  {mn@eprint@\@tempb}{\@tempb:\@tempc}{\expandafter \expandafter \csname
  mn@eprint@\@tempb\endcsname \expandafter{\@tempc}}}

\bibitem[\protect\citeauthoryear{{Acciari} et~al.,}{{Acciari}
  et~al.}{2020}]{extreme_magic}
{Acciari} V.~A.,  et~al., 2020, \apjs, 247, 16

\bibitem[\protect\citeauthoryear{{Aharonian}, {Khangulyan}  \&
  {Costamante}}{{Aharonian} et~al.}{2008}]{aharonian_khangulyan}
{Aharonian} F.~A.,  {Khangulyan} D.,   {Costamante} L.,  2008, \mnras, 387,
  1206

\bibitem[\protect\citeauthoryear{{Biteau} et~al.,}{{Biteau}
  et~al.}{2020}]{extreme_theo}
{Biteau} J.,  et~al., 2020, Nature Astronomy, 4, 124

\bibitem[\protect\citeauthoryear{{Blandford} \& {Eichler}}{{Blandford} \&
  {Eichler}}{1987}]{blandford_eichler}
{Blandford} R.,  {Eichler} D.,  1987, \physrep, 154, 1

\bibitem[\protect\citeauthoryear{{Blandford}, {Meier}  \&
  {Readhead}}{{Blandford} et~al.}{2019}]{blandford}
{Blandford} R.,  {Meier} D.,   {Readhead} A.,  2019, \araa, 57, 467

\bibitem[\protect\citeauthoryear{{Blumenthal} \& {Gould}}{{Blumenthal} \&
  {Gould}}{1970}]{blumenthal_gould}
{Blumenthal} G.~R.,  {Gould} R.~J.,  1970, Reviews of Modern Physics, 42, 237

\bibitem[\protect\citeauthoryear{{Bodo} \& {Tavecchio}}{{Bodo} \&
  {Tavecchio}}{2018}]{bodo_tavecchio}
{Bodo} G.,  {Tavecchio} F.,  2018, \aap, 609, A122

\bibitem[\protect\citeauthoryear{{B{\"o}ttcher}, {Dermer}  \&
  {Finke}}{{B{\"o}ttcher} et~al.}{2008}]{bottcher_dermer}
{B{\"o}ttcher} M.,  {Dermer} C.~D.,   {Finke} J.~D.,  2008, \apjl, 679, L9

\bibitem[\protect\citeauthoryear{{B{\"o}ttcher}, {Reimer}, {Sweeney}  \&
  {Prakash}}{{B{\"o}ttcher} et~al.}{2013}]{bottcher_reimer}
{B{\"o}ttcher} M.,  {Reimer} A.,  {Sweeney} K.,   {Prakash} A.,  2013, \apj,
  768, 54

\bibitem[\protect\citeauthoryear{{Bresci}, {Lemoine}, {Gremillet}, {Comisso},
  {Sironi}  \& {Demidem}}{{Bresci} et~al.}{2022}]{bresci_lemoine}
{Bresci} V.,  {Lemoine} M.,  {Gremillet} L.,  {Comisso} L.,  {Sironi} L.,
  {Demidem} C.,  2022, arXiv e-prints, p. arXiv:2206.08380

\bibitem[\protect\citeauthoryear{{Cerruti}, {Zech}, {Boisson}  \&
  {Inoue}}{{Cerruti} et~al.}{2015}]{cerruti}
{Cerruti} M.,  {Zech} A.,  {Boisson} C.,   {Inoue} S.,  2015, \mnras, 448, 910

\bibitem[\protect\citeauthoryear{{Chang} \& {Cooper}}{{Chang} \&
  {Cooper}}{1970}]{chang_cooper}
{Chang} J.~S.,  {Cooper} G.,  1970, Journal of Computational Physics, 6, 1

\bibitem[\protect\citeauthoryear{{Chiaberge} \& {Ghisellini}}{{Chiaberge} \&
  {Ghisellini}}{1999}]{ghisellini_chiaberge}
{Chiaberge} M.,  {Ghisellini} G.,  1999, \mnras, 306, 551

\bibitem[\protect\citeauthoryear{{Costamante} et~al.,}{{Costamante}
  et~al.}{2001}]{costamante}
{Costamante} L.,  et~al., 2001, \aap, 371, 512

\bibitem[\protect\citeauthoryear{{Costamante}, {Bonnoli}, {Tavecchio},
  {Ghisellini}, {Tagliaferri}  \& {Khangulyan}}{{Costamante}
  et~al.}{2018}]{extreme_nustar}
{Costamante} L.,  {Bonnoli} G.,  {Tavecchio} F.,  {Ghisellini} G.,
  {Tagliaferri} G.,   {Khangulyan} D.,  2018, \mnras, 477, 4257

\bibitem[\protect\citeauthoryear{{Eilek}}{{Eilek}}{1979}]{eilek}
{Eilek} J.~A.,  1979, \apj, 230, 373

\bibitem[\protect\citeauthoryear{{Essey} \& {Kusenko}}{{Essey} \&
  {Kusenko}}{2010}]{essey_kusenko}
{Essey} W.,  {Kusenko} A.,  2010, Astroparticle Physics, 33, 81

\bibitem[\protect\citeauthoryear{{Fichet de Clairfontaine}, {Meliani}, {Zech}
  \& {Hervet}}{{Fichet de Clairfontaine} et~al.}{2021}]{fichet}
{Fichet de Clairfontaine} G.,  {Meliani} Z.,  {Zech} A.,   {Hervet} O.,  2021,
  \aap, 647, A77

\bibitem[\protect\citeauthoryear{{Ghisellini}, {Tavecchio}, {Foschini},
  {Ghirlanda}, {Maraschi}  \& {Celotti}}{{Ghisellini}
  et~al.}{2010}]{ghisellini_tavecchio}
{Ghisellini} G.,  {Tavecchio} F.,  {Foschini} L.,  {Ghirlanda} G.,  {Maraschi}
  L.,   {Celotti} A.,  2010, \mnras, 402, 497

\bibitem[\protect\citeauthoryear{{Ghisellini}, {Righi}, {Costamante}  \&
  {Tavecchio}}{{Ghisellini} et~al.}{2017}]{ghisellini_righi}
{Ghisellini} G.,  {Righi} C.,  {Costamante} L.,   {Tavecchio} F.,  2017,
  \mnras, 469, 255

\bibitem[\protect\citeauthoryear{{Gourgouliatos} \&
  {Komissarov}}{{Gourgouliatos} \&
  {Komissarov}}{2018}]{gourgouliatos_komissarov}
{Gourgouliatos} K.~N.,  {Komissarov} S.~S.,  2018, Nature Astronomy, 2, 167

\bibitem[\protect\citeauthoryear{{Kakuwa}}{{Kakuwa}}{2016}]{kakuwa}
{Kakuwa} J.,  2016, \apj, 816, 24

\bibitem[\protect\citeauthoryear{{Kakuwa}, {Toma}, {Asano}, {Kusunose}  \&
  {Takahara}}{{Kakuwa} et~al.}{2015}]{kakuwa_asano}
{Kakuwa} J.,  {Toma} K.,  {Asano} K.,  {Kusunose} M.,   {Takahara} F.,  2015,
  \mnras, 449, 551

\bibitem[\protect\citeauthoryear{{Kundu}, {Vaidya}  \& {Mignone}}{{Kundu}
  et~al.}{2021}]{mignone}
{Kundu} S.,  {Vaidya} B.,   {Mignone} A.,  2021, \apj, 921, 74

\bibitem[\protect\citeauthoryear{{Larsen}, {Levermore}, {Pomraning}  \&
  {Sanderson}}{{Larsen} et~al.}{1985}]{larsen_levermore}
{Larsen} E.~W.,  {Levermore} C.~D.,  {Pomraning} G.~C.,   {Sanderson} J.~G.,
  1985, Journal of Computational Physics, 61, 359

\bibitem[\protect\citeauthoryear{{Lefa}, {Rieger}  \& {Aharonian}}{{Lefa}
  et~al.}{2011}]{lefa_rieger}
{Lefa} E.,  {Rieger} F.~M.,   {Aharonian} F.,  2011, \apj, 740, 64

\bibitem[\protect\citeauthoryear{{Matsumoto}, {Komissarov}  \&
  {Gourgouliatos}}{{Matsumoto} et~al.}{2021}]{mhd_instability}
{Matsumoto} J.,  {Komissarov} S.~S.,   {Gourgouliatos} K.~N.,  2021, \mnras,
  503, 4918

\bibitem[\protect\citeauthoryear{{Miller} \& {Roberts}}{{Miller} \&
  {Roberts}}{1995}]{miller_roberts}
{Miller} J.~A.,  {Roberts} D.~A.,  1995, \apj, 452, 912

\bibitem[\protect\citeauthoryear{{Park} \& {Petrosian}}{{Park} \&
  {Petrosian}}{1996}]{park_petrosian}
{Park} B.~T.,  {Petrosian} V.,  1996, \apjs, 103, 255

\bibitem[\protect\citeauthoryear{{Romero}, {Boettcher}, {Markoff}  \&
  {Tavecchio}}{{Romero} et~al.}{2017}]{romero}
{Romero} G.~E.,  {Boettcher} M.,  {Markoff} S.,   {Tavecchio} F.,  2017, \ssr,
  207, 5

\bibitem[\protect\citeauthoryear{{Sironi} \& {Spitkovsky}}{{Sironi} \&
  {Spitkovsky}}{2011}]{sironi_spitkovsky}
{Sironi} L.,  {Spitkovsky} A.,  2011, \apj, 726, 75

\bibitem[\protect\citeauthoryear{{Sironi}, {Petropoulou}  \&
  {Giannios}}{{Sironi} et~al.}{2015}]{sironi_petropoulou}
{Sironi} L.,  {Petropoulou} M.,   {Giannios} D.,  2015, \mnras, 450, 183

\bibitem[\protect\citeauthoryear{{Tavecchio}, {Ghisellini}, {Ghirlanda},
  {Costamante}  \& {Franceschini}}{{Tavecchio}
  et~al.}{2009}]{tavecchio_ghisellini}
{Tavecchio} F.,  {Ghisellini} G.,  {Ghirlanda} G.,  {Costamante} L.,
  {Franceschini} A.,  2009, \mnras, 399, L59

\bibitem[\protect\citeauthoryear{{Tavecchio}, {Costa}  \&
  {Sciaccaluga}}{{Tavecchio} et~al.}{2022}]{tavecchio_costa}
{Tavecchio} F.,  {Costa} A.,   {Sciaccaluga} A.,  2022, MNRAS, in press, p.
  arXiv:2207.12766

\bibitem[\protect\citeauthoryear{{Weisskopf} et~al.,}{{Weisskopf}
  et~al.}{2016}]{ixpe}
{Weisskopf} M.~C.,  et~al., 2016, in {den Herder} J.-W.~A.,  {Takahashi} T.,
  {Bautz} M.,  eds,  Society of Photo-Optical Instrumentation Engineers (SPIE)
  Conference Series Vol. 9905, Space Telescopes and Instrumentation 2016:
  Ultraviolet to Gamma Ray. p. 990517

\bibitem[\protect\citeauthoryear{{Zech} \& {Lemoine}}{{Zech} \&
  {Lemoine}}{2021}]{zech_lemoine}
{Zech} A.,  {Lemoine} M.,  2021, \aap, 654, A96

\bibitem[\protect\citeauthoryear{{Zhou} \& {Matthaeus}}{{Zhou} \&
  {Matthaeus}}{1990}]{zhou_matthaeus}
{Zhou} Y.,  {Matthaeus} W.~H.,  1990, \jgr, 95, 14881

\makeatother
\end{thebibliography}

\end{document}